\RequirePackage{ifpdf}
\ifpdf 
\documentclass[pdftex]{sigma}
\else
\documentclass{sigma}
\fi

\numberwithin{equation}{section}

\begin{document}

\allowdisplaybreaks

\renewcommand{\thefootnote}{$\star$}

\renewcommand{\PaperNumber}{111}

\FirstPageHeading

\ShortArticleName{Resolutions of Identity for Some Non-Hermitian Hamiltonians.~I}

\ArticleName{Resolutions of Identity for Some Non-Hermitian\\
Hamiltonians. I.~Exceptional Point\\ in Continuous Spectrum\footnote{This
paper is a contribution to the Proceedings of the Workshop ``Supersymmetric Quantum Mechanics and Spectral Design'' (July 18--30, 2010, Benasque, Spain). The full collection
is available at
\href{http://www.emis.de/journals/SIGMA/SUSYQM2010.html}{http://www.emis.de/journals/SIGMA/SUSYQM2010.html}}}

\Author{Alexander A. ANDRIANOV~$^{\dag\ddag}$ and Andrey V. SOKOLOV~$^\dag$}

\AuthorNameForHeading{A.A. Andrianov and A.V. Sokolov}

\Address{$^\dag$~V.A.~Fock Department of Theoretical Physics,
Sankt-Petersburg State University,\\
\hphantom{$^\dag$}~198504 St. Petersburg, Russia}
\EmailD{\href{mailto:andrianov@bo.infn.it}{andrianov@bo.infn.it}, \href{mailto:avs_avs@rambler.ru}{avs\_avs@rambler.ru}}

\Address{$^\ddag$~ICCUB, Universitat de Barcelona, 08028 Barcelona, Spain}
\EmailD{\href{mailto:andrianov@icc.ub.edu}{andrianov@icc.ub.edu}}

\ArticleDates{Received August 06, 2011, in f\/inal form November 25, 2011;  Published online December 05, 2011}

\Abstract{Resolutions of identity for certain non-Hermitian
Hamiltonians  constructed from biortho\-go\-nal sets of their
eigen- and associated functions are given for the spectral problem def\/ined on entire
axis. Non-Hermitian Hamiltonians under consideration possess the
continuous spectrum and  the following peculiarities  are
investigated:  (1)~the case when there is
an exceptional point of arbitrary multiplicity situated  on a boundary of continuous spectrum;
(2)~the case when there is an exceptional point
situated inside of continuous spectrum. The reductions of
the derived resolutions of identity under narrowing of the classes of employed
test functions are revealed. It is shown that  in the case (1) some of
associated functions included into the resolution of identity are
normalizable and some of them may be not  and in the case (2)~the bounded associated
function corresponding to the exceptional point does
not belong to the physical state space.  Spectral properties of a SUSY partner
Hamiltonian for the Hamiltonian with an exceptional point  are examined.}


\Classification{81Q60; 81R15; 47B15}

\renewcommand{\thefootnote}{\arabic{footnote}}
\setcounter{footnote}{0}

\section{Introduction}
The interest to exceptional points in
non-Hermitian quantum dynamical systems has been revoked recently
\cite{berry,ancansok06,andcansok07,sok1,klaim} although
their very notion exists  quite a time \cite{pavl,naimark,heiss,schom,hern}. Their appearance can be associated to level coalescence at complex coupling constants for initially Hermitian Hamiltonians \cite{benderwu}. If existing they play an important role
in def\/inition of energy spectra and in construction of biorthogonal
bases in Riesz spaces~\cite{Riesz}.

Whereas the appearance of exceptional points in discrete spectrum
does not give rise to any principal obstacles for building a
resolution of identity, the emergence of exceptional points  inside or on the border of continuous
spectrum makes the very construction of resolutions of identity
rather sophisticated~\cite{ancansok06}. Their correct
description in brief represents the main aim of our paper (entitled as Part I)
whereas in the subsequent paper~\cite{sokSIGMA}
(entitled as Part II) the detailed proofs of the results announced
here are presented by one of us (A.V.S.).

Let us start with the notion of exceptional point  and further on outline the structure of the present work. The
spectrum of a Hermitian Hamiltonian, in general, consists  of
continuous part and discrete points. Meanwhile the spectrum of a
non-Hermitian Hamiltonian may contain also a new type of spectral
points embedded into a continuous spectrum and/or into a discrete one, namely,
exceptional points.

The {\it exceptional point} of the spectrum of one-dimensional
Hamiltonian $h$ def\/ined on entire axis is an eigenvalue
$\lambda_0$ of this Hamiltonian for which there is a normalizable
eigenfunction $\psi_0(x)$ and also a number of associated functions
\cite{naimark} $\psi_j(x)$, $j=1,\dots, n-1$:
\[
h\psi_0=\lambda_0\psi_0,\qquad (h-\lambda_0)\psi_j=\psi_{j-1},
\qquad j=1, \ldots, n-1.
\]
For a discrete spectrum the latter ones
are typically normalizable \cite{ancansok06,andcansok07}. On the
other hand  some non-normalizable
associated functions bounded or even growing at inf\/inity may be involved in building of resolution of
identity as well. It will be proven in Part~II.  The number $n$ is an algebraic multiplicity of~$\lambda_0$. Thus, in continuous spectrum one can deal with two
types of algebraic multiplicities  (which are not necessarily equal:
their dif\/ferent types for continuous spectrum are discussed in
conclusions). If an exceptional point $\lambda_0$  belongs to a discrete part of the spectrum then~$n$ simultaneously characterizes the order of a pole at $E=\lambda_0$ of
the Green function. For an exceptional point $\lambda_0$ on the border of continuous
spectrum the Green function reveals (see Section~\ref{section2.3}) a branching point
with the pole order $2n+1$ in the variable $\sqrt{E-\lambda_0}$,
where~$n$ is a maximal number of linearly independent
eigen- and formal associated functions of~$h$ for an eigenvalue~$\lambda_0$
 in the resolution of identity. When an exceptional
point lies inside of continuous part of the spectrum the pole order
may be larger than $n$ (which has the same meaning as in the previous sentence)
that is elucidated in details in conclusions.

In this paper we build resolutions of identity for certain
non-Hermitian Hamiltonians  constructed from biorthogonal sets of
their eigen- and associated functions for the spectral problem def\/ined on entire
axis. Non-Hermitian Hamiltonians under consideration are taken with
continuous spectrum and the following peculiarities are
investigated: in Section~\ref{section2} the case when there is an exceptional point of arbitrary multiplicity
situated on a boundary of continuous spectrum; in Section~\ref{section3} the case
when there is an exceptional point inside of continuous
spectrum. In Section~\ref{section4} in conclusions the dif\/ferent ways to introduce
algebraic multiplicities are discussed and the SUSY tools
\cite{coop1,fern,abi,sukum,bagsam97,ancan,ast2,samson,ferna} in regulating them are inspected.

More specif\/ically in Sections~\ref{section2} and~\ref{section3} the reductions of the
resolutions of identity under narrowing of the classes of employed test
functions are elaborated. It is shown that the bounded associated
function in an exceptional point inside of continuous
spectrum does not belong to the physical state space (i.e.\ does not belong to the complete biorthogonal
system built from eigenfunctions of the Hamiltonian and cannot be
reproduced with the help of harmonic expansion generated by an
appropriate resolution of identity).
If an
exceptional point  lies on a boundary of continuous spectrum then
some of associated functions included into the resolution of
identity are normalizable and some of them may be not, still being
elements of a rigged Hilbert space~\cite{gelfand} and its dual one, the Gelfand triple
 generalized onto biorthogonal resolutions of identity.

\section{Resolutions of identity for model
Hamiltonians\\ with an exceptional point of arbitrary multiplicity\\ at the
bottom of continuous spectrum}\label{section2}

\subsection{Basic constructions}\label{section2.1}

Let us consider the sequence of Hamiltonians,
\[
h_n=-\partial^2+{{n(n+1)}\over{(x-z)^2}},\qquad x\in\mathbb
R,\qquad\partial\equiv{d\over{dx}}, \qquad {\rm{Im}}\,z\ne0,\qquad
n=0,1,2,\ldots,\] where $h_0$ is the Hamiltonian of a free
particle and all these Hamiltonians are ${\mathcal PT}$-symmetric \cite{bender} for the choice
${\rm{Re}}\,z=0$. One can easily check that for the
Hamiltonian $h_n$ on the energy level $E=0$  there is an
eigenfunction $\psi_{n0}(x)$ and a chain of formal associated functions
$\psi_{nl}(x)$:
\begin{gather}
h_n\psi_{n0}=0,\qquad h_n\psi_{nl}=\psi_{n,l-1},\qquad
l=1,2,3, \ldots,\nonumber\\
 \psi_{n0}(x)={{(-i)^n(2n-1)!!}\over{\sqrt{2\pi}\,(x-z)^n}},\qquad
\psi_{nl}(x)={{(-i)^n(2n-2l-1)!!}
\over{\sqrt{2\pi} (2l)!! (x-z)^{n-2l}}},\qquad l=0,1,2,\ldots,\nonumber\\
 0!!=(-1)!!=1,\qquad (-2m-1)!!={{(-1)^m}\over{(2m-1)!!}},\qquad
m=1,2,3, \ldots.\label{saf}
\end{gather} Moreover for odd $n$ the functions $\psi_{nl}(x);$ $l=0, \ldots,
[n/2]$ are normalizable  (i.e.\ belong to $L^2(\Bbb R)$) and
when $l>[n/2]$ they are non-normalizable and unboundedly growing for
$x\to\pm\infty$. For even $n$ the functions
$\psi_{nl}(x) $; $l=0, \ldots,[n/2]-1$ are normalizable, the
function
\begin{gather}
\psi_{n,n/2}(x)\equiv{{(-i)^n(n-1)!!}\over{\sqrt{2\pi}\,n!!}}
\label{asf}
\end{gather} is bounded but non-normalizable and the
functions $\psi_{nl}(x)$ for $l>n/2$ are non-normalizable and
unboundedly growing for $x\to\pm\infty$.

The Hamiltonians $h_n$, $n=0, 1, 2, \dots$ are intertwined  by the operators
\begin{gather}
q_n^\pm=\mp\partial-\chi_n(x),\!\qquad \chi_n(x)={{\psi'_{n0}(x)}\over{\psi_{n0}(x)}}
\equiv-{n\over{x-z}},\!\qquad  q_n^-\psi_{n0}=0,\!\qquad
n=0,1,2, \ldots,\!\!\!\label{int1}
\end{gather} with
the help of the chain (ladder) construction
\cite{schr,infh,abei,sukum}
\[h_nq_n^+=q_n^+h_{n-1},\qquad
q_n^-h_n=h_{n-1}q_n^-\] and \begin{gather*}h_n=q_n^+q_n^-,\qquad
h_{n-1}=q_n^-q_n^+,\qquad n=0,1,2, \ldots,\qquad
h_{-1}=-\partial^2=h_0.
\end{gather*} One easily check that
\begin{gather}q_n^-\psi_{nl}=-i\psi_{n-1,l-1},\qquad
n,l=1,2,3, \ldots .\label{qp3}
\end{gather}

As well the eigenfunctions $\psi_n(x;k)$ for continuous spectrum of the
Hamiltonian $h_n$ can be produced from the eigenfunctions
\[\psi_0(x;k)={1\over\sqrt{2\pi}} e^{ikx},\qquad k\in\Bbb
R\] for the continuous spectrum of the Hamiltonian $h_0$ of a free
particle with the help of intertwining operators (\ref{int1}):
\begin{gather} \psi_n(x;k)={1\over\sqrt{2\pi}}\left({i\over
k}\right)^nq_n^+\cdots q_1^+e^{ikx},\nonumber\\
h_n\psi_n(x;k)=k^2\psi_n(x;k), \qquad
k\in(-\infty,0)\cap(0,+\infty),\qquad
n=0,1,2, \ldots .\label{psi1}
\end{gather} For the function
$\psi_n(x;k)$ one can derive  the following explicit
representation by induction,
\begin{gather}\psi_n(x;k)={1\over\sqrt{2\pi}}\,e^{ikx}\sum\limits_{m=0}^{n}{{(n+m)!}
\over{2^mm!(n-m)!}}{i^m\over{k^m(x-z)^m}},\label{psin}
\end{gather}
where from it follows that the cofactor $(i/k)^n$ in (\ref{psi1})
provides  the asymptotic form for
\begin{gather}\psi_n(x;k)={1\over\sqrt{2\pi}} e^{ikx}\left[1+O\left({1\over
x}\right)\right], \qquad x\to\pm\infty.\label{asym}
\end{gather} As
well, using (\ref{int1}) and (\ref{psi1}), we can get the following
representations for $\psi_n(x;k)$:
\begin{gather}
\psi_n(x;k)={1\over\sqrt{2\pi}} e^{ikz}\left({d\over{dt}}-{n\over t}\right)
\left({d\over{dt}}-{{n-1}\over t}\right)\cdots\left({d\over{dt}}-{1\over
t}\right)e^t\bigg|_{t=ik(x-z)}\nonumber\\
\phantom{\psi_n(x;k)}{}
 ={e^{ikz}\over{\sqrt{2\pi} i^n(x-z)^n}}\left({\partial\over{\partial k}}
-{n\over k}\right)\left({\partial\over{\partial k}}-{{n-1}\over
k}\right)\cdots\left({\partial\over{\partial k}}-{1\over
k}\right)e^{ik(x-z)}.\label{psi2}
\end{gather} At last, it follows
from (\ref{psi1}) and (\ref{psi2}) that
\begin{gather}
\psi_n(x;k)={i\over
k} q_n^+\psi_{n-1}(x;k)\nonumber\\
\phantom{\psi_n(x;k)}{}
 ={e^{ikz}\over{i(x-z)}} \left({\partial\over{\partial
k}}-{n\over k}\right)\big[e^{-ikz}\psi_{n-1}(x;z)\big],\qquad
n=1,2,3, \ldots .\label{psi6}
\end{gather}

Let us f\/ind now the connection between $\psi_n(x;k)$ and
$\psi_{nl}(x)$, $l=0, 1, 2, \dots$. For this purpose we calculate
the following derivative, using Leibnitz formula and the relation
4.2.7.14 from \cite{prud81},
 \begin{gather}
 \lim_{k\to0}{\partial^m\over{\partial
k^m}}\big[e^{-ikz}k^n\psi_n(x;k)\big]={\partial^m\over{\partial
k^m}}\left[{1\over\sqrt{2\pi}} e^{ik(x-z)}\sum\limits_{s=0}^n{{(n+s)!}\over
{2^ss!(n-s)!}}{{i^sk^{n-s}}\over{(x-z)^s}}\right]\Big|_{k=0}\nonumber\\
 \qquad{} ={{i^{n+m}n!}\over{\sqrt{2\pi}\,2^n(x-z)^{n-m}}}\sum\limits_{j=0}^{\min\{m,n\}}
(-2)^jC^j_mC^n_{2n-j}={{i^{n+m}}\over{\sqrt{2\pi}\,(x-z)^{n-m}}}\prod\limits_{s=1}^{n}
(2s-m-1)\nonumber\\
\qquad{}
 =\begin{cases}(-1)^nm! \psi_{n,m/2}(x),&m
\text{\ is even,}\\0,&m\text{\ is odd and\ }\leqslant
2n-1,\\ \displaystyle {{i^{m-n}(m-1)!!}\over{\sqrt{2\pi} (m-2n-1)!!}}{1\over{(x-z)^{n-m}}},&m\text{\
is odd and\ }>2n-1,\end{cases}\label{limp}
\end{gather} where
$C^m_n\equiv n!/[m!(n-m)!]$ is a binomial coef\/f\/icient. Thus,
\begin{gather}
\psi_{nl}(x)={{(-1)^n}\over{(2l)!}}
\lim_{k\to0}{\partial^{2l}\over{\partial
k^{2l}}}\big[e^{-ikz}k^n\psi_n(x;k)\big],\qquad
l=0,1,2,\ldots.\label{psi8}
\end{gather}

Since the continuous spectrum of the Hamiltonian $h_n$, $n=1, 2, 3,
\dots$ coincide with $[0,+\infty)$ (see (\ref{psi1})),  the
eigenvalue $E=0$ of this Hamiltonian for the chain of functions
$\psi_{nl}(x)$, $l=0, 1, 2, \dots$ is situated at the bottom of
continuous spectrum. It will be shown in Section~\ref{section2.3} which
 of these functions are included in the resolution of
identity constructed
from eigen- and associated functions of~$h_n$.

\subsection{Biorthogonality relations}\label{section2.2}

The biorthogonality relations between functions $\psi_{nl}(x)$, $l=0,\dots, n-1$ follow from (\ref{saf}):
\begin{gather}\int_{-\infty}^{+\infty}\psi_{nl}(x)\psi_{nl'}(x)\,dx=0,
\qquad l+l'\leqslant n-1.\label{ort1}\end{gather}

The biorthogonality relations between $\psi_{n0}(x)$ and $\psi_n(x;k)$
can be derived with the help of~(\ref{int1}),~(\ref{asym}) and~(\ref{psi6}):
\begin{gather}
\int_{-\infty}^{+\infty}\psi_{n0}(x)\big[k^n\psi_{n}(x;k)\big]\,dx=
i\int_{-\infty}^{+\infty}\psi_{n0}(x)\left(-\partial+{n\over{x-z}}
\right)[k^{n-1}\psi_{n-1}(x;k)]\,dx\nonumber\\
\qquad{} =-i\psi_{n0}(x)\big[k^{n-1}\psi_{n-1}(x;k)\big]\Big|_{-\infty}^{+\infty}\nonumber\\
\qquad\quad{} +
i\int_{-\infty}^{+\infty}[q_n^-\psi_{n0}(x)]
\big[k^{n-1}\psi_{n-1}(x;k)\big]\,dx=0,\qquad
n=1,2,3,\ldots.\label{ort2}
\end{gather}

The biorthogonality relations between normalizable associated functions
$\psi_{nl}(x)$ and $\psi_n(x;k)$ can be derived in the same way  with the help of~(\ref{qp3}) by
induction,
\begin{gather}
\int_{-\infty}^{+\infty}\psi_{nl}(x)[k^n\psi_{n}(x;k)]\,dx=
i\int_{-\infty}^{+\infty}\psi_{nl}(x)\left(-\partial+{n\over{x-z}}
\right)\big[k^{n-1}\psi_{n-1}(x;k)\big]\,dx\nonumber\\
\qquad{} =-i\psi_{nl}(x)[k^{n-1}\psi_{n-1}(x;k)]\Big|_{-\infty}^{+\infty}+
i\int_{-\infty}^{+\infty}[q_n^-\psi_{nl}(x)]
\big[k^{n-1}\psi_{n-1}(x;k)\big]\,dx\nonumber\\
\qquad{} =\int_{-\infty}^{+\infty}\psi_{n-1,l-1}(x)
\big[k^{n-1}\psi_{n-1}(x;k)\big]\,dx=\cdots\nonumber\\
\qquad{} =
\int_{-\infty}^{+\infty}\psi_{n-l,0}(x)
\big[k^{n-l}\psi_{n-l}(x;k)\big]\,dx=0,\qquad l=1,\ldots,
\left[{{n-1}\over2}\right].\label{ort3}
\end{gather}

The biorthogonality relations between non-normalizable formal associated
functions $\psi_{nl}(x)$, $l=[(n+1)/2], \dots, n-1$ and
$\psi_n(x;k)$,
\begin{gather}\int_{-\infty}^{+\infty}\psi_{nl}(x)
\big[k^n\psi_n(x;k)\big]\,dx=0,\qquad
l=\left[{{n+1}\over2}\right],\ldots,n-1,\label{ort?}\end{gather} can
be derived with the help of (\ref{saf}), (\ref{psin}), the Jordan
lemma and the relation 4.2.7.17 from~\cite{prud81} as follows,
\begin{gather*}
\int_{-\infty}^{+\infty}\psi_{nl}(x)[k^n\psi_n(x;k)]\,dx\\
\qquad{} =(-i)^n{{(2n-2l-1)!!}\over{2\pi (2l)!!}}
\sum_{m=0}^{2l-n} {{i^m(n+m)!}\over{2^mm!(n-m)!}}  k^{n-m}
\int_{-\infty}^{+\infty} (x-z)^{2l-n-m}e^{ikx}\,dx\\
\qquad\quad{}
 +(-i)^n{{(2n-2l-1)!!}\over{2\pi (2l)!!}}
\sum_{m=2l-n+1}^n {{i^m(n+m)!}\over{2^mm!(n-m)!}}  k^{n-m}
\int_{-\infty}^{+\infty}
{{e^{ikx}\,dx}\over{(x-z)^{n-2l+m}}}\\
\qquad{} =(-i)^n{{(2n-2l-1)!!}\over{(2l)!!}}
\sum_{m=0}^{2l-n} {{i^m(n+m)!}\over{2^mm!(n-m)!}}  k^{n-m}
\left(-i{d\over{dk}}-z\right)^{2l-n-m}\delta(k)\\
\qquad\quad{} +{\rm{sign}}\,({{\rm{Im}}\,z}) \theta ({\rm{sign}}\,({{\rm{Im}}\,z})\,k)
(-i)^n{{(2n-2l-1)!!}\over{(2l)!!}} e^{ikz}\\
\qquad\quad{} \times\sum_{m=2l-n+1}^n {{i^{n-2l+2m}(n+m)!}\over{2^mm!(n-m)!(n-2l+m-1)!}}
k^{2n-2l-1}\\
\qquad{}=(-i)^n{{(2n-2l-1)!!}\over{(2l)!!}}
\sum_{m=0}^{2l-n} {{i^m(n+m)!}\over{2^mm!(n-m)!}}  k^{n-m}
\left(-i{d\over{dk}}-z\right)^{(n-m)-2(n-l)}\delta(k)\\
\qquad\quad{} +{\rm{sign}}\,({{\rm{Im}}\,z}) \theta ({\rm{sign}}\,({{\rm{Im}}\,z})\,k)
(-1)^l2^n{{(2n-2l-1)!!(2l+1)!!}\over{(2n)!!}}
k^{2n-2l-1} e^{ikz}\\
\qquad\quad{} \times\sum_{m=2l-n+1}^n {(-1)^m\over2^m}
C^m_nC^{2l+1}_{m+n}=0,
\end{gather*}
 where
\begin{gather*}
\theta(t)=\begin{cases}1,&t\geqslant0,\\0,&t<0.\end{cases}
\end{gather*}

The formal associated functions $\psi_{nl}(x)$, $l=n, n+1, n+2,
\dots$ are not contained in the resolutions of identity (see Section~\ref{section2.3}), but it is interesting that one can write the biorthogonality
relations for these functions with $\psi_n(x;k)$  as well,
\begin{gather}
\int_{-\infty}^{+\infty}\psi_{nl}(x)\big[e^{-ikz}k^n\psi_n(x;k)\big]\,dx\nonumber\\
\qquad{}
={{(-1)^li^n}\over{2\pi(2l)!!(2l-2n-1)!!}} e^{-ikz}
\sum_{m=0}^{n} {{i^m(n+m)!}\over{2^mm!(n-m)!}}  k^{n-m}
\int_{-\infty}^{+\infty} (x-z)^{2l-n-m}e^{ikx}\,dx\nonumber\\
\qquad{}
={{(-1)^li^n}\over{(2l)!!(2l-2n-1)!!}} e^{-ikz}
\sum_{m=0}^{n} {{i^m(n+m)!}\over{2^mm!(n-m)!}}  k^{n-m}
\left(-i{d\over{dk}}-z\right)^{2l-n-m}\delta(k)\nonumber\\
\qquad{} ={{(-1)^n}\over{(2l)!!(2l-2n-1)!!}} e^{-ikz}
\sum_{m=0}^{n} {{(-1)^m(n+m)!}\over{2^mm!(n-m)!}}  k^{n-m}
\left({d\over{dk}}-iz\right)^{2l-n-m}[e^{ikz}\delta(k)]\nonumber\\
\qquad{} ={{(-1)^n}\over{(2l)!!(2l-2n-1)!!}}
\sum_{m=0}^{n} {{(-1)^m(n+m)!}\over{2^mm!(n-m)!}}  k^{n-m}
\delta^{(2l-n-m)}(k)\nonumber\\
 \qquad{} ={{1}\over{(2l)!!(2l-2n-1)!!(2l-2n)!}} \delta^{(2l-2n)}(k)\sum_{m=0}^{n}
{{(n+m)!(2l-n-m)!}\over{2^mm!(n-m)!}}\nonumber\\
\qquad{} ={1\over{(2l-2n)!}} \delta^{(2l-2n)}(k),\qquad
l=n,n+1,n+2,\ldots,\label{ort7}
\end{gather} where we have used
(\ref{saf}), (\ref{psin}) and the relation
\begin{gather*}
\sum_{m=0}^n{{(n+m)!(s-m)!}\over{2^mm!(n-m)!}}=
\sum_{m=0}^n{{(n-m+2m)(n-1+m)!(s-m)!}\over{2^mm!(n-m)!}}\\
\qquad{} =\sum_{m=0}^{n-1}{{(n-1+m)!(s-m)!}\over{2^mm!(n-1-m)!}}
+\sum_{m=1}^n{{(n-1+m)!(s-m)!}\over{2^{m-1}(m-1)!(n-m)!}}\\
\qquad{} =\sum_{m=0}^{n-1}{{(n-1+m)!(s-m)!}\over{2^mm!(n-1-m)!}}
+\sum_{m=0}^{n-1}{{(n+m)!(s-1-m)!}\over{2^{m}m!(n-1-m)!}}\\
\qquad {} =(s+n)\sum_{m=0}^{n-1}{{(n-1+m)!(s-1-m)!}\over{2^mm!(n-1-m)!}}
=\cdots=(s+n)\cdots(s-n+2)(s-n)!\\
\qquad{}
 =(s-n-1)!!(s+n)!!,\qquad s\geqslant n.
 \end{gather*}

At last, the biorthogonality relations between eigenfunctions for
continuous spectrum of the Hamiltonian $h_n$ are proved in~\cite{sokSIGMA} and take the following form:
\begin{gather}\int_{-\infty}^{+\infty}[k^n\psi_n(x;k)]
[(k')^n\psi_n(x;-k')]\,dx=
(k')^{2n}\delta(k-k').\label{ort4}\end{gather}
Let us notice that
(\ref{ort2})--(\ref{ort?}) contain (\ref{ort1}) for $l=0, \dots,
n-1$, $l'=0$ and~(\ref{ort4}) contains~(\ref{ort1}) for $l=l'=0$
due to (\ref{psi8}). The relations~(\ref{ort7}) can be derived with the help of
dif\/ferentiation  from~(\ref{ort4}) also
in view of~(\ref{psi8}).

\subsection{Resolutions of identity}\label{section2.3}

The initial resolution of identity constructed from $\psi_n(x;k)$
holds \cite{sokSIGMA},
\begin{gather}\delta(x-x')=\int_{\cal L}\psi_n(x;k)\psi_n(x';-k)\,dk,
\label{res1}
\end{gather} where $\cal L$ is an integration path in
complex $k$ plane, obtained from the real axis by its deformation
near the point $k=0$ upwards or downwards (the direction of this
deformation is of no dif\/ference since the residue of the integrand for
the point $k=0$ is equal zero in view of (\ref{psin}) and
(\ref{limp})) and the direction of $\cal L$ is specif\/ied from
$-\infty$ to $+\infty$. This resolution of identity is valid for
test functions belonging to $CL_\gamma\equiv C^\infty_{\Bbb R}\cap
L_2(\Bbb R;(1+|x|)^\gamma)$, $\gamma>-1$ as well as for some bounded
and even slowly increasing test functions (more details are
presented in \cite{sokSIGMA}) and, in particular, for eigenfunctions
$\psi_n(x;k)$ and for the associated function~(\ref{asf}).

One can rearrange the resolution of identity
(\ref{res1}) for any $\varepsilon>0$ to the forms
\begin{gather}
\delta(x-x')=\left(\int_{-\infty}^{-\varepsilon}+
\int_\varepsilon^{+\infty}\right)\psi_n(x;k)\psi_n(x';-k)\,dk\nonumber\\
\qquad {} +\sum_{l=0}^{n-1}\left({{x'-z}\over{x-z}}\right)^{\!l}
{{\psi_{n-l-1}(x;k)\psi_{n-l}(x';-k)}\over{i(x-z)}}\Big|_{-\varepsilon}^\varepsilon
+\Big({{x'-z}\over{x-z}}\Big)^{\!n} {{\sin\varepsilon(x-x')}\over{\pi(x-x')}}
\nonumber
\\
 {} \equiv\left(\int_{-\infty}^{-\varepsilon}+
\int_\varepsilon^{+\infty}\right)\psi_n(x;k)\psi_n(x';-k)\,dk+{{\sin\varepsilon
(x-x')}\over{\pi(x-x')}}\nonumber\\
\qquad {} -{{\cos\varepsilon(x-x')}\over{2\pi\varepsilon(x-z)(x'-z)}}
\sum_{l=0}^{n-1}{{(-1/4)^l}\over{\varepsilon^{2l}(x'-z)^{2l}}}
\sum_{m=0}^{\min\{2l,n-1\}}\!C_{2l+1,m,n}{{(n\!+\!2l\!+\!1\!-\!m)!}\over{(n-1-m)!}}
\left({{x'-z}\over{x-z}}\right)^m\nonumber\\
\qquad {} +{{\sin\varepsilon(x-x')}\over{\pi(x-z)}}
\sum_{l=1}^{n-1}{{(-1/4)^l}\over{\varepsilon^{2l}(x'-z)^{2l}}}
\sum_{m=0}^{\min\{2l-1,n-1\}}\!C_{2l,m,n}{{(n+2l-m)!}\over{(n-1-m)!}}
\left({{x'-z}\over{x-z}}\right)^m,\!\!\!\!\label{res3}
\\
 C_{lmn}={1\over l}\sum_{j=0}^{m}(-1)^jC^j_l
C^{l-1}_{n-m-1+2j},\nonumber\\
 n=1,2,3,\ldots,\qquad l=1,\ldots,
2n-1,\qquad m=0,\ldots,\min\{l-1,n-1\}\nonumber
\end{gather}
 (cf.\ with (68) in
\cite{ancansok06}) and, consequently, to the form
\begin{gather}
\delta(x-x')={\lim_{\varepsilon\downarrow0}}'
\Bigg\{\left(\int_{-\infty}^{-\varepsilon}+
\int_\varepsilon^{+\infty}\right)\psi_n(x;k)\psi_n(x';-k)\,dk+{{\sin\varepsilon
(x-x')}\over{\pi(x-x')}}\nonumber\\
\qquad{} -{{\cos\varepsilon(x-x')}\over{2\pi\varepsilon(x-z)(x'-z)}}
\sum_{l=0}^{n-1} \! {{(-1/4)^l}\over{\varepsilon^{2l}(x'-z)^{2l}}}
\! \sum_{m=0}^{\min\{2l,n-1\}}\! C_{2l+1,m,n}{{(n\!+\!2l\!+\!1\!-\!m)!}\over{(n-1-m)!}}
\left({{x'-z}\over{x-z}}\right)^m\nonumber\\
\qquad{} +{{\sin\varepsilon(x-x')}\over{\pi(x-z)}}
\sum_{l=1}^{n-1}\! {{(-1/4)^l}\over{\varepsilon^{2l}(x'-z)^{2l}}}
\! \sum_{m=0}^{\min\{2l-1,n-1\}}\! C_{2l,m,n}{{(n+2l-m)!}\over{(n-1-m)!}}
\left({{x'-z}\over{x-z}}\right)^m\!\Bigg\},\!\!\!\!\label{res4}
\end{gather}
where the prime $^\prime$ at the limit symbol emphasizes that this
limit is regarded as a limit in the space of distributions (see
details in \cite{sokSIGMA}).

From (\ref{res3}) and (\ref{res4}) one can try to derive various
reduced resolutions of identity similar to the resolutions~(69) and~(70) of \cite{ancansok06}, which correspond to the partial case
$n=1$, or to the resolutions~(15) and (16) of~\cite{andcansok10}. In
particular, by virtue of Lemma 3.7 from~\cite{sokSIGMA}, for test functions from
$CL_\gamma$, $\gamma>-1$ the
resolution (\ref{res4}) can be reduced  to the form
\begin{gather}
\delta(x-x')={\lim_{\varepsilon\downarrow0}}'
\Bigg\{\left(\int_{-\infty}^{-\varepsilon}+
\int_\varepsilon^{+\infty}\right)\psi_n(x;k)\psi_n(x';-k)\,dk\nonumber\\
\qquad{} -{{\cos\varepsilon(x-x')}\over{2\pi\varepsilon(x-z)(x'-z)}}
\sum_{l=0}^{n-1}\!{{(-1/4)^l}\over{\varepsilon^{2l}(x'-z)^{2l}}}
\!\sum_{m=0}^{\min\{2l,n-1\}}\!C_{2l+1,m,n}{{(n\!+\!2l\!+\!1\!-\!m)!}\over{(n-1-m)!}}
\left({{x'-z}\over{x-z}}\right)^m\nonumber\\
\qquad{} +{{\sin\varepsilon(x-x')}\over{\pi(x-z)}}
\sum_{l=1}^{n-1}\!{{(-1/4)^l}\over{\varepsilon^{2l}(x'-z)^{2l}}}
\!\sum_{m=0}^{\min\{2l-1,n-1\}}\!C_{2l,m,n}{{(n+2l-m)!}\over{(n-1-m)!}}
\left({{x'-z}\over{x-z}}\right)^m\Bigg\}.\!\!\!\!\label{res5}
\end{gather}
More reduced resolutions of identity for the partial case $n=2$
are presented in the forthcoming Section~\ref{section2.4}.

There is another way to transform the resolution (\ref{res1}) as
well. This way was used for obtaining of the resolution of identity~(26) from \cite{andcansok10}. The integral from the right-hand part
of~(\ref{res1}) is understood \cite{sokSIGMA} as follows:
\begin{gather}\int_{\cal L}\psi_n(x;k)\psi_n(x';-k)\,dk=
\mathop{{\lim}'}_{A\to+\infty}\int_{{\cal
L}(A)}\psi_n(x;k)\psi_n(x';-k)\,dk,\label{int3}
\end{gather} where
${\cal L}(A)$ is a path in complex $k$ plane, made of the segment
$[-A,A]$ by its deformation near the point $k=0$ upwards or downwards
(the direction of this deformation is of no dif\/ference as well as in the
case with $\cal L$) and the direction of ${\cal L}(A)$ is specif\/ied
from $-A$ to $A$. When using the facts that the integral in the
right-hand part of (\ref{int3}) is a standard integral (not a
distribution) and that $k^{2n}\psi_n(x;k)\psi_n(x';-k)$ is an entire
function of $k$ (see (\ref{psin})) as well as employing  the Leibniz for\-mu\-la
and the formulae (\ref{limp}) and (\ref{psi8}), we can transform
(\ref{int3}) as follows,
\begin{gather}
\int_{\cal L}\psi_n(x;k)\psi_n(x';-k)\,dk=\mathop{{\lim}'}_{A\to+\infty}
\lim_{\varepsilon\downarrow0}\Bigg\{\left(\int_{-A}^{-\varepsilon}
+\int_\varepsilon^A\right)\psi_n(x;k)\psi_n(x';-k)\,dk\nonumber\\
\qquad\quad{} +\sum_{j=0}^{2n-1}{1\over{j!}}{{{\partial^j}\over{\partial
k^j}}\big[k^{2n}\psi_n(x;k)\psi_n(x';-k)\big]\Big|_{k=0}}\int_{{\cal
L}(\varepsilon)}{{dk}\over{k^{2n-j}}}\Bigg\}\nonumber\\
\qquad{} =\mathop{{\lim}'}_{A\to+\infty}
\lim_{\varepsilon\downarrow0}\Bigg\{\left(\int_{-A}^{-\varepsilon}
+\int_\varepsilon^A\right)\psi_n(x;k)\psi_n(x';-k)\,dk\nonumber\\
\qquad\quad{} -2(-1)^n\sum_{l=0}^{n-1}{1\over{(2n-2l-1)\varepsilon^{2n-2l-1}}}
\sum_{m=0}^l\psi_{nm}(x)\psi_{n,l-m}(x')\Bigg\}\nonumber\\
\qquad{} =\mathop{{\lim}^*}_{\!\!\!\!\varepsilon\downarrow0}
\Bigg\{\left(\int_{-\infty}^{-\varepsilon}
+\int_\varepsilon^{+\infty}\right)\psi_n(x;k)\psi_n(x';-k)\,dk\nonumber\\
\qquad\quad{} -2(-1)^n\sum_{l=0}^{n-1}{1\over{(2n-2l-1)
\varepsilon^{2n-2l-1}}}
\sum_{m=0}^l\psi_{nm}(x)\psi_{n,l-m}(x')\Bigg\},\label{int4}
\end{gather}
where the latter equality is considered as a def\/inition and the
limit $\lim\limits_{\varepsilon\downarrow0}$ is regarded as point-wise one
(not as a limit in a function space).

Let us show that  the terms outside the integral in the last line of~(\ref{int4}) can be rearranged as follows,
\begin{gather}
-2(-1)^n\sum_{l=0}^{n-1}{1\over{(2n-2l-1)
\varepsilon^{2n-2l-1}}} \sum_{m=0}^l\psi_{nm}(x)\psi_{n,l-m}(x')\nonumber\\
\qquad{} =
\sum_{l=0}^{n-1}\psi_{nl}(x;\varepsilon)\psi_{n,n-1-l}(x';\varepsilon),
\label{equ21}
\end{gather} where $\psi_{nl}(x;\varepsilon)$, $l=0,
\dots, n-1$ is the chain of eigenfunction and associated
functions (formal for $l=[(n+1)/2], \dots,n-1$) of the
Hamiltonian $h_n$ for the eigenvalue $E=0$ of the form
\begin{gather}\psi_{nl}(x;\varepsilon)=\sum_{j=0}^l\alpha_j
(\varepsilon)\psi_{n,l-j}(x), \qquad l=0,\ldots,
n-1,\label{equ22}\\
 h_n\psi_{n0}(x;\varepsilon)=0,\qquad
h_n\psi_{nl}(x;\varepsilon)=\psi_{n,l-1}(x;\varepsilon),\qquad
l=1,\ldots,n-1\nonumber
\end{gather} with $\alpha_j(\varepsilon)$, $j=0, \dots,n-1$
being unknown coef\/f\/icients which will be found below. Using
(\ref{equ21}) and (\ref{equ22}), it is easy to check that
(\ref{equ21}) is valid if\/f the coef\/f\/icients $\alpha_j(x)$ satisfy
the following system,
\begin{gather}\sum_{j=0}^l\alpha_j(\varepsilon)\alpha_{l-j}(\varepsilon)=
-{{2(-1)^n}\over{(2l+1)\varepsilon^{2l+1}}},\qquad
l=0,\ldots,n-1.\label{sys23}
\end{gather}
After the redef\/inition
\[\alpha_j(\varepsilon)={{\sqrt{2}\,i^{n+1}}
\over{\varepsilon^{2j}\sqrt{\varepsilon}}}  \beta_j,\qquad
j=0,\ldots,n-1,\]
where $\beta_j$, $j=0, \dots,n-1$ are new
unknown coef\/f\/icients, the system (\ref{sys23}) takes the form
\[
\sum_{j=0}^l\beta_j\beta_{l-j}=
{1\over{2l+1}},\qquad l=0,\ldots,n-1.
\] The general solution of the
latter system can be found in the recurrent form,
\[\beta_0=\pm1,\qquad \beta_1={1\over{6\beta_0}},\qquad \beta_l=
{1\over{2\beta_0}}\left({1\over{2l+1}}-\sum_{j=1}^{l-1}\beta_j\beta_{l-j}\right),\qquad
l=2,\ldots,n-1.\] The f\/irst terms of the sequence $\beta_j$, $j=0$,
\dots,\, $n-1$ in the case $\beta_0=1$ are the following ones,
\[\beta_0=1,\qquad\beta_1={1\over6},\qquad\beta_2={{31}\over{360}},\qquad
\beta_3={{863}\over{15120}},\qquad\beta_4={{76813}\over{1814400}},\qquad\ldots .
\]

Thus, we can choose the functions $\psi_{nl}(x;\varepsilon)$ in the
form,
\begin{gather}
\psi_{nl}(x;\varepsilon)=i^{n+1}\sqrt{2\over\varepsilon}
\sum_{j=0}^l{\beta_j\over\varepsilon^{2j}}\,\psi_{n,l-j}(x),\qquad
l=0,\ldots,n-1\label{psiep}
\end{gather}
and the resolution of
identity holds,
\begin{gather}
\delta(x-x')=\mathop{{\lim}^*}_{\!\!\!\!\varepsilon\downarrow0}
\Bigg\{\left(\int_{-\infty}^{-\varepsilon}
+\int_\varepsilon^{+\infty}\right)\psi_n(x;k)\psi_n(x';-k)\,dk\nonumber\\
\phantom{\delta(x-x')=}{}
+\sum_{l=0}^{n-1}\psi_{nl}(x;\varepsilon)
\psi_{n,n-1-l}(x';\varepsilon)\Bigg\},\label{int5}
\end{gather}
 (cf.\ with (69) from \cite{ancansok06} for the case $n=1$).
Moreover, this resolution is equivalent to (\ref{res1}), i.e.\
it is valid for all test functions for which (\ref{res1}) is valid
(cf.\ with the analogous results in Section~\ref{section3} and in~\cite{andcansok10}).

The resolution of identity (\ref{int5}) contains all $n$ functions
from the chain $\psi_{nl}(x;\varepsilon)$, $l=0, \dots,n-1$ and in this resolution
for the eigenvalue $E=0$ there are no other eigen- or associated functions of the Hamiltonian~$h_n$. The order of the
pole $k=0$ for the Green function
\begin{gather}
G_n(x,x';E)=\left[{{\pi i}\over
k}\,\psi_n(x_>;k) \psi_n(x_<;-k)\right]\Big|_{k=\sqrt{E}} ,\nonumber\\
 x_>=\max\{x,x'\}, \qquad x_<=\{x,x'\},\qquad
{\rm{Im}}\,\sqrt{E}\geqslant0,\qquad
  (h_n-E) G_n=\delta(x-x')
\label{gf2}
\end{gather} considered as a function of $k=\sqrt E$ is
equal  to $2n+1$ in view of~(\ref{psin}), and the exceptional point
$E=0$ coincides with the branch point of this Green function as a
function of~$E$. One can consider this pole of the order $2n+1$ as a
result of conf\/luence of the pole of the order $n$ for the Green
function as a function of $E=k^2$ and of the factor $k\equiv\sqrt E$
from the denominator of the Green function (see (\ref{gf2})).
Thus, the number~$n$ of  linearly independent eigen- and (formal) associated
functions of the Hamiltonian $h_n$ incorporated in the resolution of identity~(\ref{int5})
for the eigenvalue $E = 0$ (exceptional point) is equal to the
order of the ``pole'' $E=0$ for the
Green function $G_n(x,x';E)$ in the sense elucidated above or, more
rigorously, the order of the pole of $G_n(x,x';E)$ as a function of
$k=\sqrt E$ is  $2n+1$ expressed in terms of the number $n$.

Let us notice that in view of (\ref{psiep}) the functions $\psi_{nl}(x;\varepsilon)$, $l=0,\dots,n-1$  satisfy the same
biorthogonality relations from Section~\ref{section2.2} as the functions
$\psi_{nl}(x)$, $l=0, \dots,n-1$.

\subsection[Example: case $n=2$]{Example: case $\boldsymbol{n=2}$}\label{section2.4}

For the Hamiltonian
\[h_2=-\partial^2+{6\over{(x-z)^2}},\qquad x\in{\Bbb R},\qquad
{\rm{Im}}\,z\ne0, \] there are
continuous spectrum eigenfunctions
\begin{gather}
\psi_2(x;k)={1\over\sqrt{2\pi}}\left[1-{{3}\over{ik(x-z)}}-
{{3}\over{k^2(x-z)^2}}\right]e^{ikx},\nonumber\\
 h_2\psi_2(x;k)=k^2\psi_2(x;k),\qquad k\in(-\infty,0)
\cup(0,+\infty)\label{p2}
\end{gather} (see (\ref{psin})) and also  the normalizable eigenfunction $\psi_{20}(x)$ and
the bounded associated func\-tion~$\psi_{21}(x)$ on the level $E=0$,
\begin{gather*}
\psi_{20}(x)=-{{3}\over{\sqrt{2\pi}\,(x-z)^2}},
\qquad\psi_{21}(x)=-{1\over{2\sqrt{2\pi}}},\qquad
h_2\psi_{20}=0,\qquad h_2 \psi_{21}=\psi_{20},
\end{gather*}
(see (\ref{saf})
and (\ref{asf})). In this case at the
point $k=0$ there is a f\/ifth order pole in the Green function $G_2(x,x';E)$
 considered as a function of
$k=\sqrt E$ (see (\ref{gf2}) and (\ref{p2})) and thereby the exceptional point $E=0$ of the
spectrum of the Hamiltonian~$h_2$ coincides with the branch point for
the Green function as a function of~$E$.

In the case under consideration  the biorthogonality relations (\ref{ort1})--(\ref{ort?}) and
(\ref{ort4}) take  the form,
\begin{gather}\int_{-\infty}^{+\infty}\psi_{20}^2(x)\,dx=0,\qquad
\int_{-\infty}^{+\infty}\psi_{20}(x)\big[k^2\psi_2(x;k)\big]\,dx=0,\label{bo1}\\
 \int_{-\infty}^{+\infty}\psi_{20}(x)\psi_{21}(x)\,dx=0,\qquad
\int_{-\infty}^{+\infty}\psi_{21}(x)\big[k^2\psi_2(x;k)\big]\,dx=0,\label{bo2}\\
\int_{-\infty}^{+\infty}\big[k^2\psi_{2}(x;k)\big]\big[(k')^2\psi_2(x;-k')\big]\,dx=
(k')^4\delta(k-k'),\label{bo3}
\end{gather} where (\ref{bo1}) are
included in (\ref{bo3}) due to the equality
\begin{gather}\psi_{20}(x)=
\lim_{k\to0}\big[k^2\psi_2(x;k)\big]\label{bo4}\end{gather} (see
(\ref{psi8})) and (\ref{bo2}) can be derived from (\ref{bo3}) in
view of (\ref{bo4}) and of the equality{\samepage
\[\psi_{21}(x)={1\over2}
\lim_{k\to0}{\partial^{2}\over{\partial
k^{2}}}\big[e^{-ikz}k^2\psi_2(x;k)\big]
\] (see (\ref{psi8}) as
well).}

It is straightforward to check that the resolution of identity
(\ref{res3}) (which is valid for test functions from $CL_\gamma$,
$\gamma>-1$ as well as for some bounded and even slowly increasing
test functions, see  Section~\ref{section2.3}), can be rewritten in the form
\begin{gather}
\delta(x-x')=\left(\int_{-\infty}^{-\varepsilon}+
\int_{\varepsilon}^{+\infty}\right)\psi_2(x;k)\psi_2(x'; -k)\,dk
\nonumber\\
\qquad{} +\big[\psi_{20}(x;\varepsilon)\psi_{21}(x';\varepsilon)+
\psi_{21}(x;\varepsilon)\psi_{20}(x';\varepsilon)\big]+
{{\sin\varepsilon(x-x')}\over{\pi(x-x')}}+{{6\sin^2{\varepsilon\over2}(x-x')}
\over{\pi\varepsilon(x-z)(x'-z)}}\nonumber\\
\qquad{} +{{12(x-x')\sin^2
{\varepsilon\over4}(x-x')\sin{\varepsilon\over2}(x-x')}\over
{\pi\varepsilon^2(x-z)^2(x'-z)^2}}+{{3[\varepsilon(x-x')-2
\sin{\varepsilon\over2}(x-x')]^2}\over
{2\pi\varepsilon^3(x-z)^2(x'-z)^2}},\label{res9}
\end{gather}
where
the eigenfunction $\psi_{20}(x;\varepsilon)$ and the associated
function $\psi_{21}(x;\varepsilon)$ of the Hamiltonian~$h_2$ read
\begin{gather*}
\psi_{20}(x;\varepsilon)=-i\sqrt{2\over\varepsilon} \psi_{20}(x)
\equiv{{3i}\over{\sqrt{\pi\varepsilon}\,(x-z)^2}},\\
 \psi_{21}(x;\varepsilon)=-i\sqrt{2\over\varepsilon}
\left[\psi_{21}(x)+{1\over{6\varepsilon^2}}
\psi_{20}(x)\right]\equiv{i\over{2\sqrt{\pi\varepsilon}}}
\left[{1}+{1\over{\varepsilon^2(x-z)^2}}\right],\\
 h_2\psi_{20}(x;\varepsilon)=0,\qquad
h_2\psi_{21}(x;\varepsilon)=\psi_{20}(x;\varepsilon).
\end{gather*}
 The
eigenfunction $\psi_{20}(x;\varepsilon)$ and the associated function
$\psi_{21}(x;\varepsilon)$ obviously satisfy  the biorthogo\-nality
relations similar to (\ref{bo1}) and (\ref{bo2}).

It is shown in \cite{sokSIGMA} that the resolution of identity
(\ref{res9}) can be reduced: a) for test functions from $CL_\gamma$,
$\gamma>-1$ to the form
\begin{gather}
\delta(x-x')={\lim_{\varepsilon\downarrow0}}'
\Bigg\{\left(\int_{-\infty}^{-\varepsilon}+
\int_{\varepsilon}^{+\infty}\right)\psi_2(x;k)\psi_2(x';\!-k)\,dk
\nonumber\\
\qquad{} +\big[\psi_{20}(x;\varepsilon)\psi_{21}(x';\varepsilon)+
\psi_{21}(x;\varepsilon)\psi_{20}(x';\varepsilon)\big]+{{6\sin^2{\varepsilon\over2}(x-x')}
\over{\pi\varepsilon(x-z)(x'-z)}}\nonumber\\
\qquad{} +{{12(x-x')\sin^2
{\varepsilon\over4}(x-x')\sin{\varepsilon\over2}(x-x')}\over
{\pi\varepsilon^2(x-z)^2(x'-z)^2}}+{{3[\varepsilon(x-x')
-2\sin{\varepsilon\over2}(x-x')]^2}\over
{2\pi\varepsilon^3(x-z)^2(x'-z)^2}}\Bigg\}\label{res7}
\end{gather}
identical to (\ref{res5}), b) for test functions from $CL_\gamma$,
$\gamma>1$ to the form
\begin{gather}
\delta(x-x')={\lim_{\varepsilon\downarrow0}}'
\Bigg\{\left(\int_{-\infty}^{-\varepsilon}+
\int_{\varepsilon}^{+\infty}\right)\psi_2(x;k)\psi_2(x';\!-k)\,dk
\nonumber\\
\qquad{} +\big[\psi_{20}(x;\varepsilon)\psi_{21}(x';\varepsilon)+
\psi_{21}(x;\varepsilon)\psi_{20}(x';\varepsilon)\big]\nonumber\\
\qquad{} +{{12(x-x')\sin^2
{\varepsilon\over4}(x-x')\sin{\varepsilon\over2}(x-x')}\over
{\pi\varepsilon^2(x-z)^2(x'-z)^2}}
+{{3[\varepsilon(x-x')-2\sin{\varepsilon\over2}(x-x')]^2}\over
{2\pi\varepsilon^3(x-z)^2(x'-z)^2}}\Bigg\}\label{res10}
\end{gather}
and c) for test functions from $CL_\gamma$, $\gamma>3$ to the form
\begin{gather}
\delta(x-x')={\lim_{\varepsilon\downarrow0}}'
\Bigg\{\left(\int_{-\infty}^{-\varepsilon}+
\int_{\varepsilon}^{+\infty}\right)\psi_2(x;k)\psi_2(x';\!-k)\,dk
\nonumber\\
\phantom{\delta(x-x')=}{}
+\psi_{20}(x;\varepsilon)\psi_{21}(x';\varepsilon)+
\psi_{21}(x;\varepsilon)\psi_{20}(x';\varepsilon)\Bigg\}.\label{res6}
\end{gather}

The latter of these resolutions of identity seems to have a more natural form
than the previous ones, but the right-hand part of the latter resolution
cannot reproduce the {\it normalizable} eigenfunction
\[\psi_{20}(x)\not\in CL_\gamma\equiv
C_{\Bbb R}^\infty\cap L^2({\Bbb R};(1+|x|)^\gamma), \qquad
\gamma>3\] because of the biorthogonality relations. With the help
of the Jordan lemma one can check that
\begin{gather}
\lim_{\varepsilon\downarrow0}\int_{-\infty}^{+\infty}
{{12(x-x')\sin^2
{\varepsilon\over4}(x-x')\sin{\varepsilon\over2}(x-x')}\over
{\pi\varepsilon^2(x-z)^2(x'-z)^2}} \psi_{20}(x)\,dx\nonumber\\
\qquad{} =
\lim_{\varepsilon\downarrow0}\bigg\{\left[-{3\over4}  e^{\pm
i\varepsilon(z-x')/2}\mp{i\over8} \varepsilon(z-x') e^{\pm
i\varepsilon(z-x')/2}+{3\over2} e^{\pm i\varepsilon(z-x')}\right.\nonumber\\
\left.\qquad\quad{}
\pm{i\over2} \varepsilon(z-x')e^{\pm i\varepsilon(z-x')}\right]
\psi_{20}(x')\bigg\} =
{3\over4} \psi_{20}(x')\label{vych2}
\end{gather}
and
\begin{gather}
\lim_{\varepsilon\downarrow0}\int_{-\infty}^{+\infty}
{{3[\varepsilon(x-x')-2\sin{\varepsilon\over2}(x-x')]^2}\over
{2\pi\varepsilon^3(x-z)^2(x'-z)^2}}  \psi_{20}(x)\,dx  \label{vych1} \\
\qquad{} =
\lim_{\varepsilon\downarrow0}\bigg\{\left[{3\over4}  e^{\pm
i\varepsilon(z-x')/2}\pm{i\over8} \varepsilon(z-x') e^{\pm
i\varepsilon(z-x')/2}-{1\over2} e^{\pm i\varepsilon(z-x')}
\right]\psi_{20}(x')\bigg\} ={1\over4} \psi_{20}(x'),\nonumber
\end{gather}
where the upper (lower) signs correspond to the case
${\rm{Im}}\,z>0$ (${\rm{Im}}\,z<0$). Hence, just two last terms of
the resolution of identity (\ref{res10}) and the corresponding terms
in the resolutions of identity (\ref{res9}) and (\ref{res7})
give a chance to reproduce $\psi_{20}(x)$ by these
resolutions (cf.\ with the analogous results in  Section~6.1 of~\cite{ancansok06}, in  Section~2 of \cite{andcansok10} and in
Section~\ref{section3} of the present paper). It is interesting that contributions
of these terms in the resolution of identity are (see Remark~3.4 in~\cite{sokSIGMA}) singular discontinuous functionals whose supports
consist of the only element which is the inf\/inity (cf.\ with the
analogous comments in  Section~2 of~\cite{andcansok10} and in
Section~\ref{section3} of the present paper).

In the case under consideration the resolution of identity (\ref{int5}) takes
 the following form,
\begin{gather*}
\delta(x-x')={\lim_{\varepsilon\downarrow0}}^*
\Bigg\{\left(\int_{-\infty}^{-\varepsilon}+
\int_{\varepsilon}^{+\infty}\right)\psi_2(x;k)\psi_2(x'; -k)\,dk
\\
\phantom{\delta(x-x')=}{}
+\psi_{20}(x;\varepsilon)\psi_{21}(x';\varepsilon)+
\psi_{21}(x;\varepsilon)\psi_{20}(x';\varepsilon)\Bigg\},
\end{gather*} (cf.\
with (\ref{res6})) and it is valid for test
functions from $CL_\gamma$, $\gamma>-1$ as well as for some bounded
and even slowly increasing test functions (see  Section~\ref{section2.3}).

\section{Resolutions of identity for the model Hamiltonian\\
with exceptional point inside of the continuous spectrum}\label{section3}

For the Hamiltonian
\begin{gather*}
h=-\partial^2+16\alpha^2{{\alpha(x-z)\sin
2\alpha x+2\cos^2\alpha x}\over{[\sin2\alpha x+ 2\alpha(x-z)]^2}},\\
 x\in\mathbb R, \qquad
\partial\equiv{d\over{dx}},\qquad \alpha>0,\qquad {\rm{Im}}\,z\ne0
\end{gather*}
there are \cite{ancansok06} continuous spectrum eigenfunctions
\begin{gather}
 \psi(x;k)={1\over\sqrt{2\pi}}\left[1+{{ik}\over{k^2-\alpha^2}}
{{W'(x)}\over{W(x)}}-{1\over{2(k^2-\alpha^2)}}{{W''(x)}\over{W(x)}}
\right]e^{ikx},\nonumber\\
W(x)= \sin2\alpha x+2\alpha(x-z),\nonumber\\
 h\psi(x;k)=k^2\psi(x;k),\qquad
k\in(-\infty,-\alpha)\cup(-\alpha,\alpha)\cup(\alpha,+\infty) .
\label{psik}
\end{gather}
As well for the level $E=\alpha^2$ there is the normalizable eigenfunction
$\psi_{0}(x)$ and the bounded associated function\footnote{There is a
misprint in the normalization of $\psi_0(x)$ and $\psi_1(x)$ in
\cite{ancansok06}.} $\psi_{1}(x)$,
\begin{gather}
 \psi_{0}(x)={{(2\alpha)^{3/2}\cos\alpha x}\over{\sin2\alpha
x+2\alpha(x-z)}}, \qquad\psi_{1}(x)={{{2\alpha}{{(x-z)}}\sin\alpha
x+\cos\alpha x} \over{\sqrt{2\alpha}\,[\sin2\alpha
x+2\alpha(x-z)]}},\nonumber\\
 \psi_{1}(x)={i\over{2\sqrt{2\alpha}}}\big[e^{-i\alpha x}-e^{i\alpha
x}\big]+O\left({1\over x}\right),\qquad
x\to\pm\infty,\label{saf1}
\end{gather} such that
\[
h\psi_{0}=\alpha^2\psi_{0},\qquad\big(h-\alpha^2\big) \psi_{1}=\psi_{0}.\]
The exceptional point $E=\alpha^2$ is a pole for the Green function
\begin{gather*}
G(x,x';E)=\left[{{\pi i}\over
k}\,\psi_n(x_>;k) \psi_n(x_<;-k)\right]\Big|_{k=\sqrt{E}} ,\\
 x_>=\max\{x,x'\}, \qquad x_<=\{x,x'\},\qquad
{\rm{Im}}\,\sqrt{E}\geqslant0,\qquad (h-E)\,G=\delta(x-x').
\end{gather*}
 This
pole is a pole of second order, it is replicated on both
sides of the cut
$E>0$ and there are no other poles for~$G(x,x';E)$.

One can show \cite{ancansok06} that the eigenfunctions and the
associated function of $h$ obey the biorthogo\-nality relations,
\begin{gather}\int_{-\infty}^{+\infty}\psi_{0}^2(x)\,dx=0,\qquad
\int_{-\infty}^{+\infty}\psi_{0}(x)\big[\big(k^2-\alpha^2\big)\psi(x;k)\big]\,dx=0,
\label{ort11}\\
\int_{-\infty}^{+\infty}\psi_{0}(x)\psi_{1}(x)\,dx=0, \qquad
\int_{-\infty}^{+\infty}\psi_1(x)\big[\big(k^2-\alpha^2\big)\psi(x;k)\big]\,dx=0,
\label{ort11'}\\
\int_{-\infty}^{+\infty}[(k^2-\alpha^2)\psi(x;k)]
\big[\big((k')^2-\alpha^2\big)\psi(x;-k')\big]\,dx=\big((k')^2-\alpha^2\big)^2\delta(k-k'),
\label{ort12}
\end{gather}
where (\ref{ort11}) are included in
(\ref{ort12}) due to the equality \begin{gather}\psi_0(x)=\mp
i\sqrt{\pi\over\alpha}\,\lim_{k\to\pm\alpha}
\big[\big(k^2-\alpha^2\big)\psi(x;k)\big]\label{psi0k}
\end{gather} and
(\ref{ort11'}) follow from (\ref{ort12}) in view of (\ref{psi0k})
and of the equality
\[\psi_1(x)=\pm i\sqrt{\pi\over\alpha}\,\lim_{k\to\mp\alpha}
\left\{{1\over{2k}} {\partial\over{\partial k}}
\big[\big(k^2-\alpha^2\big)\big]\psi(x;k)\right\}-{{1\mp 2i\alpha
z}\over{4\alpha^2}} \psi_0(x).
\]

The resolution of identity constructed from $\psi(x;k)$ holds \cite{sokSIGMA},
\begin{gather}\delta(x-x')=\int_{\cal L}\psi(x;k)\psi(x';-k)\,dk,
\label{res01}\end{gather}
where $\cal L$ is an integration path in
complex $k$ plane, obtained from the real axis by its simultaneous
deformation near the points $k=-\alpha$ and $k=\alpha$ upwards or
downwards (the direction of this deformation is of no dif\/ference since for the points $k=-\alpha$ and
$k=\alpha$ the
sum of residues of the integrand  is equal to zero). The direction of $\cal L$ is
specif\/ied from $-\infty$ to $+\infty$. This resolution of identity
is valid for test functions belonging to $CL_\gamma\equiv
C^\infty_{\Bbb R}\cap L_2(\Bbb R;(1+|x|)^\gamma)$, $\gamma>-1$ as
well as for some bounded and even slowly increasing test functions
 and, in particular,
for eigenfunc\-tions~$\psi(x;k)$ and for the associated
function~$\psi_1(x)$.

One can rearrange \cite{sokSIGMA} the resolution of identity
(\ref{res01}) for any $\varepsilon\in(0,\alpha)$ to the form
\begin{gather}
\delta(x-x')=\left(\int_{-\infty}^{-\alpha-\varepsilon}+
\int_{-\alpha+\varepsilon}^{\alpha-\varepsilon}+
\int_{\alpha+\varepsilon}^{+\infty}\right) \psi(x;k)\psi(x';-k)\,dk\nonumber\\
\phantom{\delta(x-x')=}{}
+{2\over\pi} \cos\alpha(x-x'){{\sin\varepsilon(x-x')}
\over{x-x'}}-{1\over{\pi\alpha}}\,\psi_0(x)\psi_0(x')
\left[{1\over\varepsilon}\left[1-2\sin^2{\varepsilon\over2} (x-x')\right]\right.\nonumber\\
\left.\phantom{\delta(x-x')=}{}
-{\varepsilon\over{4\alpha^2-\varepsilon^2}} \cos2\alpha(x-x')
\cos\varepsilon(x-x')-{{2\alpha}\over{4\alpha^2-\varepsilon^2}} \sin2\alpha(x-x')
\sin\varepsilon(x-x')\!\right]\nonumber\\
\phantom{\delta(x-x')=}{}
-{1\over{\pi}} [\psi_0(x)\psi_1(x')+\psi_1(x)\psi_0(x')]
\int_{2\alpha-\varepsilon}^{2\alpha+\varepsilon}\cos
t(x-x') {{dt}\over t}\label{res13}
\end{gather}
and, consequently,
to the form
\begin{gather}
\delta(x-x')={\lim_{\varepsilon\downarrow0}}'
\Bigg\{\left(\int_{-\infty}^{-\alpha-\varepsilon}+
\int_{-\alpha+\varepsilon}^{\alpha-\varepsilon}+
\int_{\alpha+\varepsilon}^{+\infty}\right) \psi(x;k)\psi(x';-k)\,dk\nonumber\\
\phantom{\delta(x-x')=}{}
 +{2\over\pi} \cos\alpha(x-x'){{\sin\varepsilon(x-x')}
\over{x-x'}}-{1\over{\pi\alpha}}\,\psi_0(x)\psi_0(x')
\left[{1\over\varepsilon}\left[1-2\sin^2{\varepsilon\over2} (x-x')\right]\right.\nonumber\\
\left. \phantom{\delta(x-x')=}{}
-{\varepsilon\over{4\alpha^2\!-\varepsilon^2}} \cos2\alpha(x-x')
\cos\varepsilon(x-x')-{{2\alpha}\over{4\alpha^2\!-\varepsilon^2}} \sin2\alpha(x-x')
\sin\varepsilon(x-x')\!\right]\nonumber\\
\phantom{\delta(x-x')=}{}
 -{1\over{\pi}} [\psi_0(x)\psi_1(x')+\psi_1(x)\psi_0(x')]
\int_{2\alpha-\varepsilon}^{2\alpha+\varepsilon}\cos
t(x-x') {{dt}\over t}\bigg\},\label{res14}
\end{gather} where the
prime $^\prime$ at the limit symbol emphasizes that this limit is
regarded as a limit in the space of distributions.

One can reduce \cite{sokSIGMA} the resolutions of identity
(\ref{res13}) and (\ref{res14}) for test functions from $CL_\gamma$,
$\gamma>-1$ to the form
\begin{gather}
\delta(x-x')={\lim_{\varepsilon\downarrow0}}'
\Bigg\{\left(\int_{-\infty}^{-\alpha-\varepsilon}+
\int_{-\alpha+\varepsilon}^{\alpha-\varepsilon}+
\int_{\alpha+\varepsilon}^{+\infty}\right) \psi(x;k)\psi(x';-k)\,dk\nonumber\\
\phantom{\delta(x-x')=}{}
-{1\over{\pi\varepsilon\alpha}}\left[1-
2\sin^2{\varepsilon\over2} (x-x')\right]\psi_{0}(x)\psi_{0}(x')\Bigg\}
\label{res11}
\end{gather} and for test functions from $CL_\gamma$,
$\gamma>1$ to a more simple form
\begin{gather}
\delta(x-x')={\lim_{\varepsilon\downarrow0}}'
\Bigg\{\!\left(\int_{-\infty}^{-\alpha-\varepsilon}\!\!+
\int_{-\alpha+\varepsilon}^{\alpha-\varepsilon}\!\!+
\int_{\alpha+\varepsilon}^{+\infty}\right)\! \psi(x;k)\psi(x';-k)\,dk
-{1\over{\pi\varepsilon\alpha}} \psi_{0}(x)\psi_{0}(x')\Bigg\}.\!\!\!\!\!
\label{res12}
\end{gather}

The latter of these resolutions seems to have a more natural form
than the previous ones, but it cannot reproduce the {\it
normalizable} eigenfunction
\[\psi_{0}(x)\not\in
CL_\gamma\equiv C_{\Bbb R}^\infty\cap L^2({\Bbb R};(1+|x|)^\gamma),
\qquad \gamma>1\] because of the biorthogonality relations. With the
help of~(\ref{saf1}), Lemma~4.8 from~\cite{sokSIGMA} and the Jordan
lemma one can check that
\begin{gather}
\lim_{\varepsilon\downarrow0}\int_{-\infty}^{+\infty}
\left[{2\over{\pi\varepsilon\alpha}} \sin^2{\varepsilon\over2} (x-x')
\psi_{0}(x)\psi_{0}(x')\right] \psi_0(x)\,dx\nonumber\\
\qquad{} =\lim_{\varepsilon\downarrow0}\Bigg\{\psi_{0}(x')\int_{-\infty}^{+\infty}
\left[{2\over{\pi\varepsilon\alpha}} \sin^2{\varepsilon\over2}(x-x')\right]\nonumber\\
\qquad\quad{} \times\left[ 2\alpha{{\cos^2\alpha x}\over{(x-z)^2}}-{{\sin2\alpha
x [\sin2\alpha
x+4\alpha(x-z)]}\over{4\alpha^2(x-z)^2}} \psi^2_{0}(x)\right] dx\Bigg\}\nonumber\\
\qquad{} =\lim_{\varepsilon\downarrow0}\Bigg\{\psi_{0}(x')\int_{-\infty}^{+\infty}
\left[{2\over{\pi\varepsilon\alpha}} \sin^2{\varepsilon\over2} (x-x')\right]\left[
2\alpha{{\cos^2\alpha x}\over{(x-z)^2}}\right] dx\Bigg\} \label{vosp12}\\
\qquad{} =\lim_{\varepsilon\downarrow0}\Bigg\{
\left[ e^{\pm
i\varepsilon(z-x')}-4{\alpha\over\varepsilon}\,\sin^2{\varepsilon\over
2} (z-x')\pm ie^{\pm 2i\alpha z}\sin\varepsilon(z-x')\right]
\psi_{0}(x')\Bigg\} =\psi_0(x'),\nonumber
\end{gather} where the upper
(lower) signs correspond to the case ${\rm{Im}}\,z>0$
(${\rm{Im}}\,z<0$). Hence, just the term
\begin{gather}{2\over{\pi\varepsilon\alpha}}
 \sin^2{\varepsilon\over2} (x-x') \psi_{0}(x)\psi_{0}(x')
\label{term}
\end{gather} in the resolutions of identity
(\ref{res13})--(\ref{res11}) gives an opportunity to
reproduce $\psi_0(x)$ by these resolutions (cf.\ with the analogous
results in Section~6.1 of \cite{ancansok06}, in  Section~2 of
\cite{andcansok10} and in  Section~\ref{section2.4} of the present paper). It is
interesting that the contribution of the term (\ref{term}) in the
resolutions of identity (\ref{res13})--(\ref{res11}) is  a singular discontinuous functional (see
Remark~4.2 in~\cite{sokSIGMA})
which support consists of the only element~-- the inf\/inity (cf.\
with the analogous comments in  Section~2 of \cite{andcansok10} and
in  Section~\ref{section2.4} of the present paper).

There is another way to transform the resolution (\ref{res01}) as
well. This way was used for obtaining of the resolutions of identity~(26) in \cite{andcansok10} and (\ref{int5}) in   Section~\ref{section2.3}. The
integral in the right-hand part of (\ref{res01}) is understood
as follows,
\begin{gather}\int_{\cal L}\psi(x;k)\psi(x';-k)\,dk=
\mathop{{\lim}'}_{A\to+\infty}\int_{{\cal
L}(A)}\psi(x;k)\psi(x';-k)\,dk,\label{int03}
\end{gather} where
${\cal L}(A)$ is an integration path in complex $k$ plane, obtained
from the segment $[-A,A]$ by its simultaneous deformation near the
points $k=-\alpha$ and $k=\alpha$ upwards or downwards (the direction
of this deformation is of no dif\/ference as well as in the case with $\cal
L$) and the direction of ${\cal L}(A)$ is specif\/ied from~$-A$ to
$A$. Using
\renewcommand{\labelenumi}{\rm{(\theenumi)}}
\begin{enumerate}\itemsep=0pt

\item the fact that the integral in the right-hand part of
(\ref{int03}) is a standard integral (not a~distribution);

\item the fact that $(k\mp\alpha)^2\psi(x;k)\psi(x';-k)$ is a~holomorphic function of $k$ in a neighborhood of $k=\pm\alpha$ (see
(\ref{psik}));

\item the Leibniz for\-mu\-la and the formulae (\ref{saf1});

\item the notation ${\cal L}(k_0;\varepsilon)$ with f\/ixed $k_0\in\Bbb
R$ and $\varepsilon>0$ for the path in complex $k$ plane def\/ined~by
\[
k=k_0+\varepsilon[\cos(\pi-\vartheta)\pm i\sin(\pi-\vartheta)],
\qquad 0\leqslant\vartheta\leqslant\pi,
\] where the upper (lower)
sign corresponds to the case of upper (lower) deformations in $\cal
L$ and the direction of ${\cal L}(k_0;\varepsilon)$ is specif\/ied
from $\vartheta=0$ to $\vartheta=\pi$;
\end{enumerate}
we can transform (\ref{int03}) as follows,
\begin{gather}
\int_{\cal L}\psi(x;k)\psi(x';-k)\,dk=\mathop{{\lim}'}_{A\to+\infty}
\lim_{\varepsilon\downarrow0}\Bigg\{\left(\int_{-A}^{-\alpha-\varepsilon}
+\int_{-\alpha+\varepsilon}^{\alpha-\varepsilon}
+\int_{\alpha+\varepsilon}^A\right)\psi(x;k)\psi(x';-k)\,dk\nonumber\\
\qquad\quad{} +\sum_{j=0}^{1}{1\over{j!}}{{\partial^j}\over{\partial
k^j}} \big[(k+\alpha)^2\psi(x;k)\psi_n(x';-k)\big]\Big|_{k=-\alpha}\int_{{\cal
L}(-\alpha;\varepsilon)}{{dk}\over{(k+\alpha)^{2-j}}}\nonumber\\
\qquad\quad{} +\sum_{j=0}^{1}{1\over{j!}}{{\partial^j}\over{\partial
k^j}} \big[(k-\alpha)^2\psi(x;k)\psi_n(x';-k)\big]\Big|_{k=\alpha}\int_{{\cal
L}(\alpha;\varepsilon)}{{dk}\over{(k-\alpha)^{2-j}}}\Bigg\}\nonumber\\
\qquad{} =\mathop{{\lim}'}_{A\to+\infty}
\lim_{\varepsilon\downarrow0}\Bigg\{\left(\int_{-A}^{-\alpha-\varepsilon}
+\int_{-\alpha+\varepsilon}^{\alpha-\varepsilon}
+\int_{\alpha+\varepsilon}^A\right)\psi(x;k)\psi(x';-k)\,dk\nonumber\\
\qquad\quad{}
+{1\over{4\pi\alpha}} \psi_0(x)\psi_0(x')\left[\int_{{\cal
L}(-\alpha;\varepsilon)}{{dk}\over{(k+\alpha)^2}}+\int_{{\cal
L}(\alpha;\varepsilon)}{{dk}\over{(k-\alpha)^2}}\right]\nonumber\\
\qquad\quad{}
+{1\over{2\pi}} [\psi_0(x)\psi_1(x')+\psi_1(x)\psi_0(x')]
\left[\int_{{\cal
L}(\alpha;\varepsilon)}{{dk}\over{k-\alpha}}-\int_{{\cal
L}(-\alpha;\varepsilon)}{{dk}\over{k+\alpha}}\right]\Bigg\}\nonumber\\
\qquad{}
=\mathop{{\lim}'}_{A\to+\infty}
\lim_{\varepsilon\downarrow0}\Bigg\{\left(\int_{-A}^{-\alpha-\varepsilon}
+\int_{-\alpha+\varepsilon}^{\alpha-\varepsilon}
+\int_{\alpha+\varepsilon}^A\right)\psi(x;k)\psi(x';-k)\,dk
-{1\over{\pi\varepsilon\alpha}} \psi_0(x)\psi_0(x')\Bigg\}\nonumber\\
\qquad{}=\mathop{{\lim}^*}_{\!\!\!\!\varepsilon\downarrow0}
\Bigg\{\left(\int_{-\infty}^{-\alpha-\varepsilon}
+\int_{-\alpha+\varepsilon}^{\alpha-\varepsilon}
+\int_{\alpha+\varepsilon}^{+\infty}\right)\psi(x;k)\psi(x';-k)\,dk
-{1\over{\pi\varepsilon\alpha}} \psi_0(x)\psi_0(x')\Bigg\},\!\!\!\!\!
\label{int04}
\end{gather} where the latter equality is considered
as a def\/inition for $\lim^*$ and the limit $\lim_{\varepsilon\downarrow0}$ is
regarded as a pointwise one (not as a limit in a function space). The
resolution of identity (\ref{int04}) is equivalent to (\ref{res01}),
i.e.\ it is valid for all test functions for which
(\ref{res01}) is valid (cf.\ with (\ref{res12}) and with the
similar results in  Section~2 of~\cite{andcansok10} and in
Section~\ref{section2.3} of the present paper).

Let us note that the associated function
$\psi_1(x)$ does not appear in the derived resolutions of
identity and is not expandable with the help of the resolution \eqref{res12}. Thereby, this associated function does not belong to the
physical state space (rigged Hilbert space).

Let us notice also that the number (equal to 1) of linearly
independent eigen- and (formal) associated functions of the
Hamiltonian $h$ for the eigenvalue $E=\alpha^2$ included into the
resolutions of identity (\ref{res12}) and (\ref{int04}) is less than
the order (equal to 2) of the pole $E=\alpha^2$ of the Green
function $G(x,x';E)$.

\section{Conclusions: indexes of exceptional points and SUSY}\label{section4}

We remark that, in general, one can introduce, at least, three  dif\/ferent number indexes
of exceptional point $E=\lambda_0$ of a Hamiltonian $h$:
\renewcommand{\labelenumi}{\rm{(\theenumi)}}
\begin{enumerate}\itemsep=0pt
\item $n_1(\lambda_0)$ to be a  maximal number of linearly independent
normalizable eigenfunctions and associated functions of $h$ for the
eigenvalue $E=\lambda_0$;

\item $n_2(\lambda_0)$ to be a  maximal number of linearly independent
eigenfunctions and formal associated functions of $h$ for the
eigenvalue $E=\lambda_0$ appeared in the resolution of identity
constructed from biorthogonal set of eigenfunctions and associated
functions of $h$;

\item $n_3(\lambda_0)$ to be an order of the pole in the point
$E=\lambda_0$ for the Green function for $h$ as a function of $E$ (in
the case with the Hamiltonian $h=h_n$ (see  Section~\ref{section2}), where the
exceptional point $E=\lambda_0=0$ coincides with the branch point of
the Green function; it is natural to assume (see  Section~\ref{section2.3}) that
$n_3(0)=n$).
\end{enumerate}
It can be proven (by methods of \cite{naimark}, see as well the example
in  Section~5.1 of \cite{ancansok06}) that for an exceptional point  outside of
continuous spectrum all these indexes are identical,
\[n_1(\lambda_0)=n_2(\lambda_0)=n_3(\lambda_0)\] and represent the
algebraic multiplicity of the eigenvalue $\lambda_0$. In the cases
where an exceptional point is situated on the border of continuous
spectrum or inside of it these indexes may be
dif\/ferent. For example, in the case of an exceptional
point $E=0$  at the bottom of continuous spectrum of the
Hamiltonian $h_n$ in  Section~\ref{section2},
\[n_1(0)=\left[{{n+1}\over2}\right],\qquad n_2(0)=n,\qquad
n_3(0)=n\qquad\Rightarrow\qquad n_1(0)\leqslant n_2(0)=n_3(0)\] and
in the case of an exceptional point $E=\alpha^2$ inside of  continuous spectrum of the Hamiltonian $h$  in Section~\ref{section3},
\[n_1\big(\alpha^2\big)=1,\qquad n_2\big(\alpha^2\big)=1,\qquad n_3\big(\alpha^2\big)=2
\qquad\Rightarrow\qquad n_1\big(\alpha^2\big)=n_2\big(\alpha^2\big)<n_3\big(\alpha^2\big).\]
Thus, one can consider these indexes as dif\/ferent generalizations of
the notion of algebraic multiplicity.

One can use SUSY technique \cite{coop1,fern,abi,sukum,bagsam97,ancan,ast2} in order to
regulate the algebraic multiplicity (in any sense mentioned above) of
an exceptional point in the spectrum of a SUSY partner Hamiltonian
with respect to the order of this exceptional point in the spectrum
of a~given Hamiltonian (originally proposed in \cite{samson} and elaborated in details
in \cite{andcansok07,sok1}):
\renewcommand{\labelenumi}{\rm{(\theenumi)}}
\begin{enumerate}\itemsep=0pt

\item in order to increase the multiplicity of an exceptional point
$\lambda_0$ in the spectrum of a SUSY partner Hamiltonian with
respect to its order in the spectrum of a given Hamiltonian, one
must take  a formal eigenfunction (and
a chain of formal associated functions)  for the spectral value $\lambda_0$ of the latter Hamiltonian as transformation function(s)
 which tends (tend) to inf\/inity
for $x\to\pm\infty$;

\item in order to decrease the multiplicity of an exceptional point
$\lambda_0$ in the spectrum of a SUSY partner Hamiltonian with
respect to its multiplicity in the spectrum of a given Hamiltonian,
one must take  a normalizable eigenfunction (and a
chain of normalizable associated functions) of the latter Hamiltonian for
eigenvalue $\lambda_0$ as transformation function(s).
\end{enumerate}

These statements can be clarif\/ied by the following simple example.
For the eigenvalue $E=\lambda_0=0$ of the Hamiltonian $h_n$, $n=1,
2, 3, \dots $ in  Section~\ref{section2} there is a chain (\ref{saf}) of the
eigenfunction $\psi_{n0}(x)$ and  associated functions
$\psi_{nl}(x)$, $l=1, \dots,  [(n-1)/2]$. As well one can
check that for the spectral value $E=0$ of the Hamiltonian
$h_n$, $n=0, 1, 2, \dots$  there is a chain of the formal eigenfunction
$\varphi_{n0}(x)$ and formal associated functions $\varphi_{nl}(x)$,
\begin{gather*}
\varphi_{n0}(x)=(x-z)^{n+1},\qquad\varphi_{nl}(x)={{(-1)^l(2n+1)!!}
\over{(2l)!!(2n+2l+1)!!}} (x-z)^{n+2l+1},\\
 h_n\varphi_{n0}=0,\qquad h_n\varphi_{nl}=\varphi_{n,l-1},\qquad
l=1,2,3,\ldots,
\end{gather*}
 which tend to inf\/inity for $x\to\pm\infty$. Thus,
it can be easily found that a) if to use $\varphi_{nl}(x)$, $l=0,
\dots, m$ as transformation functions for $h_n$, then the resulting
Hamiltonian is $h_{n+m+1}$ with the exceptional point $E=0$ of
larger algebraic multiplicity (except for the case with the indexes~$n_1(0)$, $m=0$ and odd~$n$, where the indexes~$n_1(0)$ for~$h_n$
and $h_{n+m+1}\equiv h_{n+1}$ are equal), and b) if to use
$\psi_{nl}(x)$, $l=0, \dots, m\leqslant[(n-1)/2]$ as
transformation functions for $h_n$, then the resulting Hamiltonian
is $h_{n-m-1}$ with the exceptional point $E=0$ of smaller algebraic
multiplicity (except for the case with indexes $n_1(0)$, $m=0$ and
even $n$, where the indexes $n_1(0)$ for $h_n$
and $h_{n-m-1}\equiv h_{n-1}$ are equal).

\subsection*{Acknowledgments} This work was supported by
Grant RFBR 09-01-00145-a and by the SPbSU project 11.0.64.2010. The
work of A.A.\ was also  supported by grants   2009SGR502,
FPA2007-66665 and by the Consolider-Ingenio 2010 Program CPAN
(CSD2007-00042).

\pdfbookmark[1]{References}{ref}
\LastPageEnding

\end{document}